# Sustainable and Resilient Systems for Intergenerational Justice

Sahar Zandi[1]

[1]*College of Architecture, Texas A&M University, College Station, TX 77843*

**Abstract:**

Rawls' theory of justice aims at fairness. He does not only think of justice between exiting parties in existing society, but he also thinks of it between generations – intergenerational justice problem. Rawls' solution to this problem is saving principle and he says that we are responsible for being just with the next generations. Wolf thinks of our responsibility for future generations as a kind of financial debt that we have to pay. He also develops the meaning of "saving" and says that it is not restricted to the monetary one. Wolf extends the definition of "saving" such that includes investment on behalf of the next generations as well.

In this paper, I want to extend the meaning of "saving" to "using the resources for sustainable and resilient systems." By referring to "the problem of time," I show that our decision on behalf of the next generations will be rational only if we entirely use the natural resources and wealth to produce sustainable and resilient systems.

## 1.The problem of intergenerational justice

Justice is the significant concept of Rawls' philosophy, and "justice as fairness" is his theory of justice. He aims at proposing some grounds for social justice in institutions of a liberal society. And his theory of justice considers both freedom and equality of the citizens. However, he does not only focuses on justice between the parties in existing society but also thinks of justice between generations. He says: "persons in different generations have duties and obligations to one another just as contemporaries do." He dedicated section 44 to the intergenerational justice problem, and he begins it by referring to the question of justice between generations and says that this issue has a crucial role for every ethical theory. He also states that that:

> "These comments about how to specify the social minimum have led us to the problem of justice between generations. Finding a just savings principle is one aspect of this question." (Rawls 1999)

He explains that we can not determine the social minimum, regardless of the important principle of intergeneration justice. He doesn't think of justice as a temporal requirement which only considers the current generation, but he also thinks of as s historical approach, which leads to some considerations between generations.

There are two main questions here. First, are we, i.e., the individuals of the current generation, are responsible for the justice between us and the next generations? Rawls responds that every generation is responsible for justice with the next generations, but not vice versa (due to the fact that it is not possible.) Secondly, how can we be just regearing the next generations? In the next section, I will elaborate on Rawls' response to this question.

## 2.Rawls solution: the principle of saving

Rawls' response to the second question will be started by referring to the original position. He states that we can not think of intergenerational justice through a democratic approach because the only existing generation is the current generations, and we don't know anything about the next generations. So, we have to think of the original position by using the veil of ignorance. That is, we can consider that all the individuals in all generations will be present in a thought experience, and imagine that it is not determined that we are in the current generation or not. Then, the result of our consideration will be so closed to a fair and correct idea on the intergenerational justice problem. Rawls says that:

> "Since no one knows to which generation he belongs, the question is viewed from the standpoint of each and a fair accommodation is expressed by the principle adopted. All generations are virtually represented in the original position since the same principle would always be chosen. An ideally democratic decision will result, one that is fairly adjusted to the claims of each generation and therefore satisfying the precept that what touches all concerns all." (Rawls 1999)

Accordingly, we can decide democratically through this thought experiment in the original position. Then, he states that utilitarianism doesn't work here well because "the utilitarian principle may lead to an extremely high rate of saving, which imposes excessive hardships on earlier generations." The expected utility of the next generations will definitely be larger than any present sacrifice. However, it is not fair, and it seems that utilitarianism is still less justified in the intergeneration justice problem.

### 2.1. Saving principle

Rawls' solution is the saving principle. He proposes that we would consider all generations in the original position and then invoking the veil of ignorance, we can think of the saving principle as a solution to this problem. He states that:

> "The just savings principle can be regarded as an understanding between generations to carry their fair share of the burden of realizing and preserving a just society." (Rawls 1999)

That is, every generation has to partially save the natural recourses and wealth and transfer it to the next generations. In this way, every generation will benefit from the last generation and will transfer the saving to the next generations, and so the justice between generations will be realized. Definitely, the first generation can not benefit from the outcome of the other generations, but it is the nature of life, and this restriction is inevitable. But, the first generation can play a role in making this intergenerational justice.

### 3. Wolf's formulation of intergenerational justice: financial debt

Rawls proposes the saving principle as a general solution for the intergenerational justice problem; however, he didn't elaborate on the details of this idea in different contexts and situations.(Afroogh 2019 & 2020) Wolf, in "Justice and Intergenerational Debt," states that

> "We can think of environmental debts in the same way as financial debts, and that this will help us to understand our most important obligations of intergenerational justice." (Wolf 2008)

He clarifies that we can not decide on behalf of the next generations; however, we have to think about the effect of our actions on them. He gives an example of the huge amount of debt of the USA government and the implications of this debt for the next generation. And says that

> "It is unjust for present generations to pass on a debt burden to their successors except where those burdens are fully compensated " (Wolf 2008)

He also adds that:

> "Many of the costs we pass on to later generations are non-monetary, but they have precisely the same structure as a monetary debt: Where our present actions damage or degrade the natural environment, we pass on a burden that can be measured in terms of the rate at which the environment can recover from our assaults. The rate of recovery translates to a measure of the cost we pass on, since" (Wolf 2008)

Then he discusses some major approaches in order to make intergeneration justice, and he proposes that:

> "As individual persons, our saving and consumption rates are usually planned around the life-cycle changes we expect to live through. But as nations, or as a global society, we might plan for a longer time horizon." (Wolf 2008)

Accordingly, he defends using the natural resources in a sustainable way such that the next generations can benefit from their outcomes. That is, in addition to saving money and natural resource, we can make integrational justice by investing in a sustainable way for the next generations. As it is clear, Wolf's approach toward intergenerational justice is different from the traditional idea of saving money and natural resource for the next generation. He prescribes that the current generation can invest the savings for the next generation as well. In what follows, I want to go beyond this and state that it is not rational to save money or natural recourse if we, the current generation, can produce sustainable and resilient systems.

## 4. Sustainability and community resilience for intergeneration justice

I agree with Rawls' main idea in saving principle, which says we have to consider the rights of all individuals in all generations in the original position. It seems to me that making intergenerational justice would not be possible without caring about the next generations. However, I think we don't need to follow a traditional sense of "saving" – that is, saving some amount of money or natural resources for the next generations and don't use them. We need some new interception of "saving" in the Rawls' saving principle.

I agree with Wolf's idea of investing in a sustainable way for the next generation. I also think that all resources are not material ones, and we also need to care about non-material resources and opportunities, like environmental resources, scientific development, political development, etc. However, I want to extend the meaning of "saving". By referring to "the problem of time," I shall show that the rational way for "saving" the resources for the next generations is "fully using" them in order to produce sustainable and resilient systems, which will guarantee the most possible outcomes for the next generation.

### 4.1 The problem of "time"

The fast growth of science, economy, technology, etc., shows that the problem of "time" is significant in human life development. Therefore, the delay in using or investing in natural recourse and money will cause some huge costs for current human lives as well as the next generations. It is not optimal just to save, in a traditional sense, some amount of money or recourses for the next generations, due to the fact that we are ignoring the significance of time in human life development. It is not rational to delay investment the natural recourse for the next generations, and it would be a kind of betrayal if we ignore it because this delay will cause a huge cost for the lives of the next generations.

Put it in other words, by saving, in a traditional sense, for the next generations, actually we are deciding on behalf of the next generations for their wealth. It looks a prisoner dilemma regarding the independent decision-making process (HajiAliAkbari and Esmaeili, 2021; Esmaeili and Hajialiakbari, 2021). However, we will not be rational representatives for them if we just put their wealth aside, without increasing and changing it to the better sustainable and resilient ones, which are pretty possible regarding the long time at work. A rational decision on behalf of the next generations is to "fully use" their money and natural resources in a sustainable and resilient way, such that both persevere the original money and recourse and increase it rationally in the long period of time. In this way, we will prevent the cost of delay in investment in the traditional sense of "saving" and will maximize the interests of the investments as well. In what follows, I will elaborate on the two concepts of a sustainable and resilient system, which I think would be the most important constituents for a new interpretation of "saving" for intergenerational justice (Afroogh 2021).

## 4.2 Sustainable and resilient systems

Since the 1980s, the concept of sustainability has been used to focus on preserving and maintaining human life, particularly for future generations focusing on justice and fairness (Mebratu 1998). Afterward, the concept of resilience was proposed as a necessary feature of an appropriate sustainable engineering system, referring to the reflective capability feature of a system (Meerow, Newell & Stults, 2016) & (Marchese, 2018).

Sustainably is an adjective for all kinds of systems and designs in human civilization. Sustainable systems and designs refer to the human system, including engineering, economical, cultural, etc., systems that preserve the original recourse and transfer it to the next generations in addition to the new achievements and outcome of these systems. For example, a sustainable engineering system is a system that uses the existing resources and adds to it, and transfer the original resources in addition to the new outcomes (of the system) to the next generation. So, we don't need to save anything for the next generation. We use all the resources, as much as possible, yet in a very smart and rational way such that we add to the original values of these resources and then transfer it to the next generation. Nowadays, a sustainable system is a huge literature in many engineering and empirical science departments, and interdisciplinary research will lead us to find the relation between those systems and social justice.

Community resilience has been recently defined in terms of recovering the capability of stakeholders, which are affected by disaster and natural hazards (Doorn, Gardoni, & Murphy, 2018). It refers to a property of sustainable systems. That is, every sustainable system ought to be communally resilient. Community resilience is actually a response to a problem for sustainable systems. The problem is that there is some huge disaster in human life, like floods, earthquakes, etc. and they will destroy all humans systems. So, we can not transfer all these systems and outcomes to the next generations, and so it is not justice. Therefore, in response to this problem, and to guarantee justice between generations, we invoke the community resilience concept. It briefly says that we have to design every institution and system such that it can overcome the natural disasters and repair and reform the system automatically to come back to the original and previous status of the system before the disaster. Therefore, by creating and using resilient and sustainable systems, we can guarantee intergenerational justice.

An inclusive community resilience aims at justice and is committed to bringing about human well-being, which refers to the well-being of each and all human individuals irrespective of their gender, sex, race, education, financial status, etc. Studies in vulnerability assessment of the communities have revealed that the socially vulnerable population experience higher hardship from natural disasters(Coleman et al. 2019). This has been shown to be rooted in individuals' higher exposure to the threats and their lower ability to tolerate the negative impacts(Lindell et al. 2006). The social inequality in the societal impacts of natural disasters suggests that current approaches have failed to meet the needs of the affected communities. This highlights the importance of incorporating justice in designing resilient infrastructure systems. Therefore, a system would be resilient only if it aims at justice and equally consider each and all individuals.

## 5. Conclusion

The traditional meaning of "saving" doesn't suffice for intergenerational justice. We have to consider the new concepts of the sustainable and resilient systems instead of the traditional sense of "saving" in order to guarantee intergenerational justice.

Regarding the problem of time, our decisions on behalf of the next generations (about their rights on natural recourse and wealth) would be rational only if we fully use the natural resources and wealth to produce sustainable and resilient systems. We have to extend the meaning of "saving" in Rawls' saving principle to the complete usage of the resources in producing sustainable and resilient systems.

**References:**

- Afroogh, Saleh (2019). Contextual Reason and Rationality. Master's thesis, Texas A&M University. Available electronically from http : / /hdl .handle .net /1969 .1 /186349. 10.13140/RG.2.2.21462.06726
- Afroogh, Saleh. "A Contextualist Decision Theory." arXiv preprint arXiv:2101.08914 (2021).
- Afroogh, Saleh. "De Dicto Cognitive Reason Contextualism."(2020)
- John Rawls, *A Theory of Justice*, revised edition (Harvard, 1999)
- HajiAliAkbari, Mahdi, and Shahin Esmaeili. "Prisoner Dilemma in maximization constrained: the rationality of cooperation." arXiv preprint arXiv:2102.03644 (2021).
- Shahin Esmaeili, and Mahdi HajiAliAkbari,"Robert Nozick on Prisoner's Dilemma" (2021).
- Wolf, Clark (2008): Justice and Intergenerational Debt. In: Intergenerational Justice Review 1/2008 (Vol. 8). pp. 13- 17.
- Mebratu, D., 1998. Sustainability and sustainable development: historical and conceptual review. Environmental Impact Assessment Review 18, 493 – 520.
- Meerow, S., Newell, J. P., & Stults, M. (2016). Defining urban resilience: A review. Landscape and Urban Planning, 147, 38–49. http://doi.org/10.1016/j. landurbplan.2015.11.011

- Marchese, D.; Reynolds, E.; Bates, M.E.; Morgan, H.; Clark, S.S.; Linkov, I. Resilience and sustainability: Similarities and differences in environmental management applications. Sci. Total Environ. 2018, 613–614, 1275–1283. [CrossRef] [PubMed]
- Coleman, N., Esmalian, A. & Mostafavi, A. Equitable Resilience in Infrastructure Systems: Empirical Assessment of Disparities in Hardship Experiences of Vulnerable Populations during Service Disruptions. *Nat. Hazards Rev.* (2019).
- Lindell M, Perry R, Prater C, Nicholson W (2006) Fundamentals of Emergency Management. 479
- Doorn, N., Gardoni, P., & Murphy, C. (2018). A multidisciplinary definition and evaluation of resilience: The role of social justice in defining resilience. *Sustainable and Resilient Infrastructure*, 1–12.  (Rawls 1999.)
- Afroogh, Saleh. (2020), De Dicto Cognitive Reason Contextualism.10.13140/RG.2.2.14888.80649